\documentclass[10pt,conference]{IEEEtran}

\IEEEoverridecommandlockouts

\usepackage{cite}
\usepackage{amsmath,amssymb,amsfonts}
\usepackage{algorithmic}
\usepackage{algorithm}
\usepackage{comment}
\usepackage{graphicx}
\usepackage{caption}
\usepackage{subcaption}
\usepackage{graphics}
\usepackage{textcomp}
\usepackage{xcolor}
\usepackage[left=0.68in, right=0.673in, top=0.78in, bottom=1.1in]{geometry}

\begin{document}
\title{Data-Driven Radio Environment Map Estimation Using Graph Neural Networks}

\author{\IEEEauthorblockN{Ali Shibli\IEEEauthorrefmark{2}, 
Tahar Zanouda}\
\IEEEauthorblockA{Global AI Accelerator, Ericsson, Sweden}\
\texttt{\{ali.a.shibli, tahar.zanouda\}@ericsson.com}\\
\IEEEauthorblockA{\IEEEauthorrefmark{2}KTH Royal Institute of Technology\\\texttt{shibli@kth.se}}
}


\maketitle




\begin{abstract}
\label{section:abstract}
Radio Environment Maps (REMs) are crucial for numerous applications in Telecom. The construction of accurate Radio Environment Maps (REMs) has become an important and challenging topic in recent decades. In this paper, we present a method to estimate REMs using Graph Neural Networks. This approach utilizes both physical cell information and sparse geo-located signal strength measurements to estimate REMs. The method first divides and encodes mobile network coverage areas into a graph. Then, it inputs sparse geo-located signal strength measurements, characterized by Reference Signal Received Power (RSRP) and Reference Signal Received Quality (RSRQ) metrics,  into a Graph Neural Network Model to estimate REMs. The proposed architecture inherits the advantages of a Graph Neural Network to capture the spatial dependencies of network-wide coverage in contrast with network Radio Access Network node locations and spatial proximity of known measurements.

Extensive experiments on real-world data demonstrate that the proposed model outperforms four other machine learning models, and shows its ability to generalize across different geographical regions.

\end{abstract}

\section{Introduction}
\label{section:intro}

Telecom networks are becoming increasingly complex and heterogeneous. Mobile networks are constantly evolving, requiring network operators to continuously introduce changes to network parameters and expand the network infrastructure. As the complexity of the operating network increases, mobile operators face new challenges to continuously assess the network and handle upgrade plans. Traditionally, mobile operators used Radio Environment Maps (REMs) to estimate network service in various coverage areas. REM refers to multi-domain radio signal quality information (typically the received signal quality) in various geographical areas. REMs are estimated using measurements taken from the network. The data that characterizes the interaction between user equipment (UE) and the network can be captured in several ways. For instance, \textit{Drive Test} is a method of measuring and assessing the coverage, capacity, and Quality of Service (QoS) of a mobile radio network by collecting \textit{minimization of drive test} measurements, which are standardized by 3GPP \cite{lin2023artificial}. These measurements consist of UE information, field measurements, radio measurements, and location information. \textit{Signal strength measurements} are used to interpolate and estimate radio maps. While the spatial preciseness and the high observation frequency of signal strength measurements are advantageous, the sparsity of sampling process of collecting radio measurements remains a challenge. 

The accurate REM estimation is an important task for numerous applications \cite{dare2023radio} in the telecom domain such as network configuration and parameters optimization, spectrum estimation, coverage optimization, finding optimal locations for new cells, and proactive resource management.

In this work, we refer to REMs of received Reference Signal Received Power (RSRP) in 4G/5G  networks. These REMs are constructed using sparse signal strength measurements. Radio signal quality is estimated for areas and locations that have not been measured. The aim of this paper is to construct REMs using sparse measurements that were obtained while sampling the network signal quality. Previous methods propose various techniques to address these challenges. These methods, although powerful, require domain knowledge (e.g., calculate path loss) and are often computationally costly. 

In light of these limitations, we propose a new method using Graph Neural Networks. The main contributions of this work are summarized as follows:
\begin{enumerate}
    \item A suitable model architecture for REM estimation using Graph Convolution Networks (GCNs) that is competitive with alternative solutions.
    \item Comparative analysis with state-of-the-art models: Providing a thorough evaluation of the proposed GNN model against current machine learning models such as XGBoost algorithm, known for its efficiency in structured/tabular data \cite{chen2016xgboost}, deep learning model based on Fully Connected Network (FCN), and TabNet \cite{arik2021tabnet}, a recent innovation in deep learning for tabular data, renowned for its interpretability and performance.
    \item Comprehensive data integration and data fusion combining diverse data types, including physical cell information, geo-located signal strength measurements, and geographic information.
\end{enumerate}

The paper is organized as follows. Section II reviews the state-of-the-art. Data is discussed in section III. Section IV mathematically formulates the problem, and outlines the proposed method. Section V provides details on evaluation metrics, experimental results, and analysis are introduced in Section VI and a conclusion to conclude the paper.

\section{Previous work}
\label{section:literature_review}

In recent years, there has been a growing body of research \cite{li2018recent} \cite{dare2023radio} that focuses on estimating Radio Environment Maps. Radio Environment Map estimation techniques either relied on physical and statistical
propagation models to describe signal quality propagation properties (i.e., statistical approaches) or employ a variety of deep learning algorithms (i.e., data-driven approaches). 
Traditional and statistical approaches are criticized for relying on domain knowledge and assumptions which can limit their applications to real-world scenarios. Methods such \cite{zhu20213gpp} are based on ray tracing and stochastic radio propagation model. Such models require a detailed mapping system that emulates the target real-world environment. However, the built-up infrastructure evolves over time and weather conditions change frequently which makes the process of modeling the environment a time-consuming process.

Recently, machine learning approaches have become increasingly popular due to the prevalence of data. Alimpertis et al. \cite{alimpertis2019city} used Random Forest to generate predictive signal strength maps based on measurements and other information such as device location.

Rufaida et al. \cite{rufaida2020construction} conducted a comparative analysis of several models: KNN, SVM, and two decision tree-based models (XGBoost and LightGBM), and showed that tree-based models outperform other models. The work was only applied to wireless local area networks. 

Anjinappa et al \cite{anjinappa2021coverage} proposed a data-driven unsupervised learning algorithm named uniform manifold approximation and projection (UMAP) with a focus on millimeter-wave networks. Millimeter waves have distinctive properties such its sensitivity to physical obstacles such as buildings and trees. The planning of millimeter-wave (mmWave) bands in 5G networks usually undergo different processes. 

Levie et al. \cite{levie2021radiounet} proposed a Deep Learning method to estimate path loss from a transmitter location to any point in a flat domain. The results of the paper showed that the path loss function was estimated in an urban environment with high accuracy and low computational complexity.

Teganya et al. \cite{teganya2021deep} proposed a Deep Learning architecture based on a fully convolutional deep completion autoencoder to estimate radio maps. The numerical experiments with two datasets
showed that the method can attain good RMSE.

Thrane et al. \cite{thrane2020deep} presented a model-aided deep learning approach for path loss prediction. In their work, they augmented radio data with rich and unconventional information about the site, e.g. satellite photos, to provide more accurate and flexible models. Similarly, authors \cite{zhang2020cellular} used top-view geographical images for Radio propagation modeling.

Chaves-Villota et al. \cite{chaves2023deeprem} proposed Deep Learning method, coined as \textit{DeepREM}, to estimate REMs in urban scenarios using combined U-Net and Conditional Generative Adversarial Network (CGAN) DL architectures.

The approaches, although powerful, struggle to handle data sparsity and do not incorporate the spatial proximity of coverage areas.

\section{Data}
\label{section:data}

\begin{figure}[H]
    \centering
    \includegraphics[width = 0.99\linewidth]{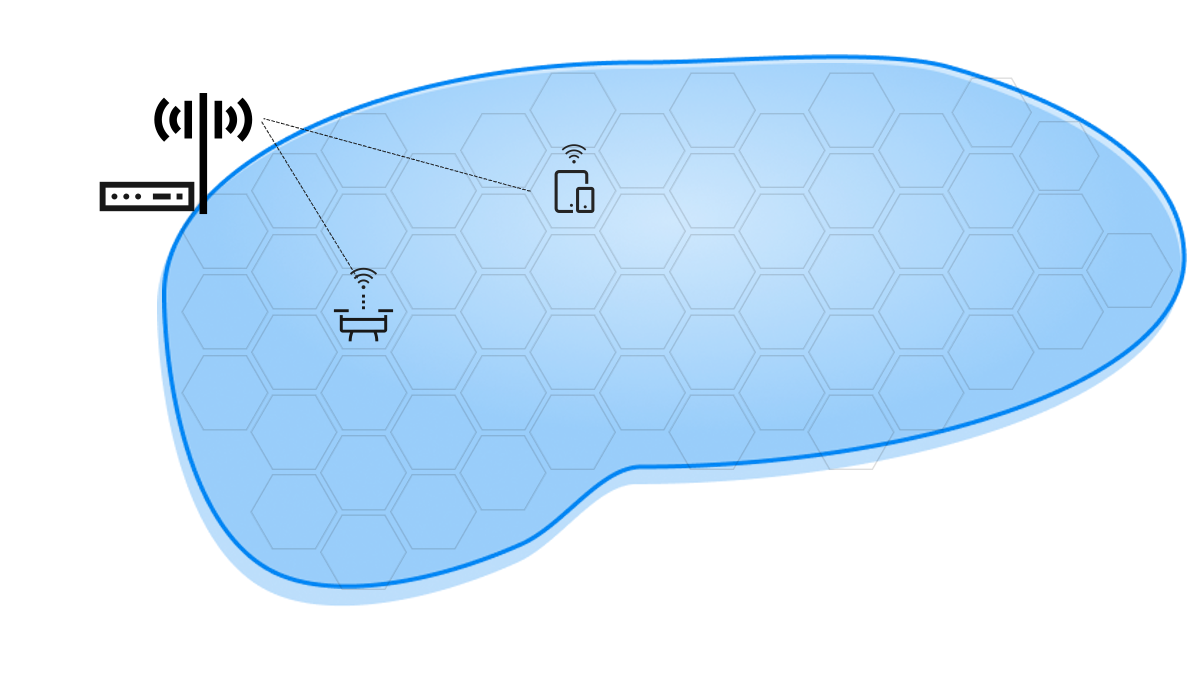}
    \caption{A figure showing a coverage area, divided in hexagons, where UEs receive signal from cells}
    \label{fig:teaser}
\end{figure}

A telecom operational network consists of interconnected Radio nodes (e.g., gNB, eNB, etc) that provide coverage to end-users. In each coverage area, User equipment (UE) communicates via Radio nodes. The coverage area can be divided into tiles (hexagons) as shown in \ref{fig:teaser} where UEs receive a signal from the cell. The interaction between UE and the Radio node can be captured in several ways. With the rising ubiquity of network data, mobile operators can sample geo-signal strength measurements to estimate signal quality. Geo-located signal strength measurements can be sampled from different devices connecting with the network to sense signal strength and network coverage. Sparse signal strength measurements contains information about signal quality in some regions in the city.

In our work, we use a dataset that spans \textit{8 weeks} and is collected from \textit{2 cities}. Each data point in the dataset encompasses a set of attributes such as global cell identifier, signal strength measurement values, timestamp of observation, and the longitude and latitude information of signal measurements. The dataset does not contain any information about the devices or users. The dataset covers two cities and includes +1000 deployed Radio Nodes. Such datasets can be collected using Minimization of Drive Tests (MDT) feature to assess performance of the network. 

\section{Method}
\label{section:method}
This section details the model architectures and the method for generating REM.

\subsection{Problem Definition}

Given a city $C$, with scattered signal measurements. This work aims to utilize the collected measurements to predict signal quality in data-insufficient tiles to facilitate the task of generating coverage maps.

\subsection{Method Overview}
We propose a Graph Neural Network (GNN)-based approach to estimate REMs.
The main steps are described in \ref{alg:main_algo}.

\begin{algorithm}
\begin{algorithmic}[1]
    \FOR{each time interval $t = 1, \ldots, T$}

    \STATE Obtain live network inventory data, for each Radio Node $\mathcal{C}_{t}$.
    \STATE Obtain plurality of geo-located signal strength measurements $M$.
    \STATE Map each geo-located signal strength measurements $M_{t}$ to a Radio Nodes $\mathcal{C}_{t}$ using Global Cell ID.
    \STATE Generate tiles $\mathcal{R}_{t}$ for a geographic area of interest using hierarchical geospatial indexing system, H3 \cite{sahr2003geodesic}. 
    \STATE Determine adjacency matrix, by finding neighboring tiles for each tile.
    \STATE Construct the graph. The city is represented as a temporal graph, consisting of tiles $\mathcal{R}_{t}$ and edges $\mathcal{E}_{t}$ connecting the tile
    \STATE Prepare ML model features: 
    \FOR{each tile $\mathcal{R}_{t}$ in the city, }
    \STATE - Calculate distances and bearing values between geo-located signal strength measurements $M_{t}$, that belong to the same tile, and the associated Radio Node  $\mathcal{C}_{t}$.  
    \STATE - Find Radio Nodes $\mathcal{C}_{t}$ associated with each tile and extract frequency configuration parameters.
    \STATE - Extract geospatial features for each tile $\mathcal{R}_{t}$.
    \STATE - Extract historical measurements characterized by Reference Signal Received Power (RSRP) and Reference Signal Received Quality (RSRQ) metrics associated with each tile.
    \ENDFOR
    
    
    \STATE Pre\-process ML model features:
    \\- Apply one-hot encoding on categorical variables.
    \\- Normalize continuous values.
    \STATE Train GNN model using the model features prepared in previous steps, and the adjacency matrix extracted from the city graph.
    \STATE Estimate RSRP and RSRQ measurements for each tile $\mathcal{R}_{t}$ in the city using a trained GNN model, where the sparse measurements are missing.
    \ENDFOR
    
\end{algorithmic}
\caption{Main steps of the method.}
\label{alg:main_algo}
\end{algorithm}   
                
\subsection{Overview of the process}
The pipeline consists of the following  steps:

\subsubsection{Data pre-processing}

In the first step of the pipeline, during the pre-processing of the data, the data is processed according to the required input format of the network architecture. The process begins with the collection of diverse datasets:
\begin{enumerate}
    \item Sampled signal strength measurements.
    \item Network infrastructure data including Radio node location.
    \item Spatial data encompasses the geographical locations of network sites.
\end{enumerate}

These datasets undergo feature engineering to extract and structure the data into a format amenable for processing by the GNN. Two GNN models are evaluated:
(1) node regression model, where the GNN predicts a continuous RSRP score, and (2) node classification, where it categorizes RSRP values into four distinct classes— \textit{Very weak}, \textit{Weak}, \textit{Average}, and \textit{Good} according to Table \ref{tab:rsrp_classification}. This dual approach allows for a comprehensive understanding of signal quality across different dimensions, enhancing the accuracy and utility of the coverage maps generated.

The datasets are then preprocessed as described in the following steps:
\begin{enumerate}
    \item  \textit{Min-max} scale for continuous values in each set to the range \textit{[0,1]}.
    \item One-hot encoding for categorical values.
\end{enumerate} 

\begin{table}[htbp]
    \centering
    \begin{tabular}{ll}
    \hline
    \textbf{RSRP (in dBm)} & \textbf{Classification Label} \\
    \hline
    Less than -120 & Very weak \\
    Between -120 and -106 & Weak \\
    Between -105 and -90 & Average \\
    Greater than -90 & Good \\
    \hline
    \end{tabular}
    \caption{Classification of RSRP values.}
    \label{tab:rsrp_classification}
\end{table}

\subsubsection{Coverage Map Representation using a Temporal Graph}

In this work, we partition the city into $K$ equally-sized regions (or tiles), each represented in a hexagon shape. Each region is represented by a unique geographic identifier corresponding to a region on the Earth. In our work, we use the H3 index from Sahr et al. \cite{sahr2003geodesic}. H3 is a framework comprising of a global grid system that is suitable for analyzing large spatial datasets, by partitioning areas of the Earth into identifiable grid tiles such as hexagons. The resolution/hexagon size reflects the size of homogeneous hexagons used to divide the earth. The choice of the hexagon size can be fine-tuned during the training process. H3 divides regions into approximately equal areas regardless of location on the planet. It is important to consider equal areas when developing a solution at a global scale. Dividing geographical regions into squares can result in biased results between northern or southern parts of the globe and regions close to the equator. With a hierarchical hexagon grid, we can control the granularity of the region for each use case and easily find the right resolution for each use case.
Each sample in the dataset (the signal strength measurements) is mapped into the corresponding tile, represented by a geographic identifier corresponding to a region on the Earth. The H3 method is used to map (latitude, longitude) to spatial hexagons and generate the geo index (geo hash). It is worth mentioning that there are other indexing systems (e.g., S2) that divide geographic regions into rectangular squares. However, H3 keeps both hexagon area and shape distortion to a minimum together, compared to square and rectangular grids. 

We treat each hexagon as a Region $R$. Each Region $R$ is characterized by a set of features associated with a vector $X_r$.
We structure the city as a temporal graph, consisting of regions $\mathcal{R}$ and edges $\mathcal{E}$. It is represented as follows:
$(\mathcal{G}_{\text{C}}^t)_{t=1}^{T} = (\{\mathcal{R}^t, \mathcal{E}\})_{t=1}^{T}$, 
where $\mathcal{R}^t=\{x_1^t, \ldots, x_k^t\}$ represents the vertex features corresponding to the multivariate time series $(\mathbf{x}^t)_{t}$ at time step $t$, and $(\mathcal{E})_t=\mathcal{E}$ is the static edge set of the region graph at time step $t$.
For each hexagon tile, we determine its neighbors in the H3 grid system. The number of neighboring levels considered can be a hyper-parameter. For instance, we benchmarked k-level graphs (kth outer level circle of neighbors) as well.
\begin{figure}[H]
    \centering
    \includegraphics[width = 0.99\linewidth]{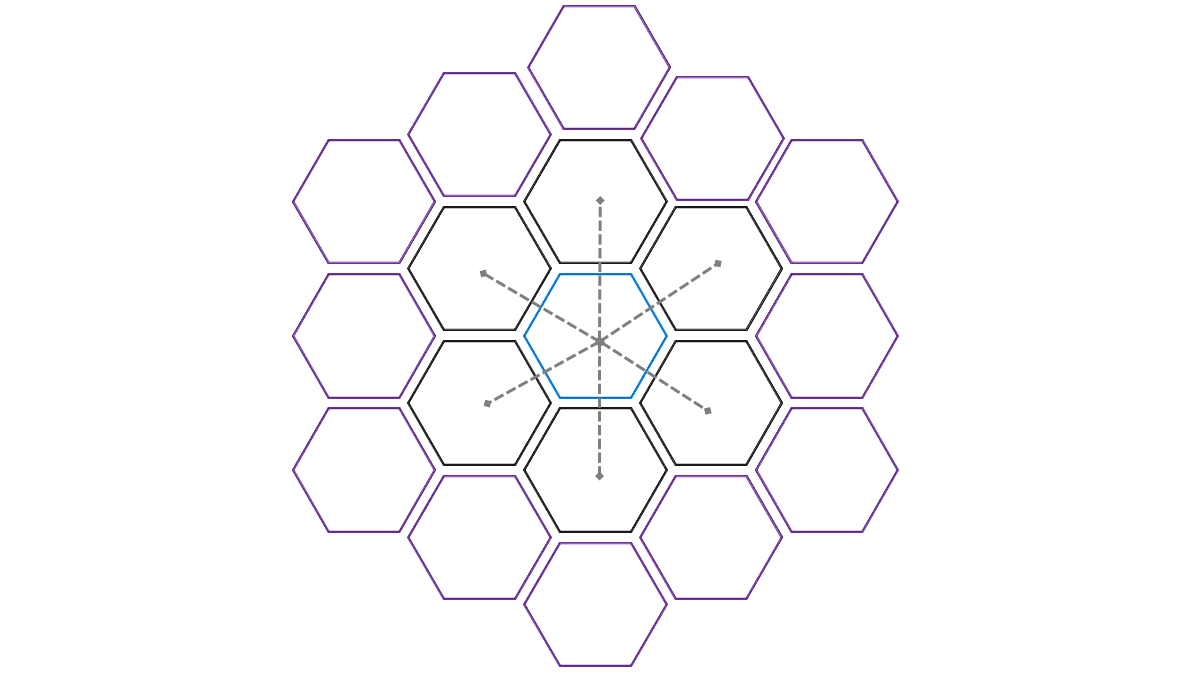}
    \caption{Representing Coverage Areas using H3 Hexagons}
    \label{fig:model}
\end{figure}

\subsubsection{Feature Engineering}

Once the measurements are grouped by each cell in the network, we perform the following data processing tasks to prepare a set of features for the model:
\begin{itemize}
    \item (1) Cell Parameters encompassing attributes like Channelbandwidth (data transmission capacity) and Earfcndl (central downlink frequency channel number), Radio Access Technology (4G/5G).
    \item (2) Historical geo-located signal strength measurements including RSRQ (Reference Signal Received Quality), RSRP (Reference Signal Received Power), RSSI (Received Signal Strength Indicator), and SINR/SNR (Signal-to-noise ratio of the given signal).
    \begin{itemize}
    \item (2.1) Index measurement location: we use the H3 method \cite{sahr2003geodesic} to index measurements (measurement latitude, measurement longitude) to spatial hexagon and generate geo index (geo hash).
    \item (2.2) Calculate the Bearing azimuth between the measurement location and cell location. Bearing azimuth, in this context, refers to the angular direction of the signal measurement’s location corresponding to the cell location. The values can vary between 0 and 360.
    \item (2.3) Calculate the geographic distance between the sample location and cell location.
    \item (2.4) Extract historical signal strength values such as RSRQ, RSRP, RSSI, and SINR/SNR.
    \item (2.5) Extract statistics on device brand usage.
    \end{itemize}
    \item (3) Spatial context values displaying geographical attributes like terrain type, and functional area type.
\end{itemize}

\subsubsection{Coverage Map Estimation Model using Graph Neural Networks}
\begin{figure}[H]
    \centering
    \includegraphics[width = 1\linewidth]{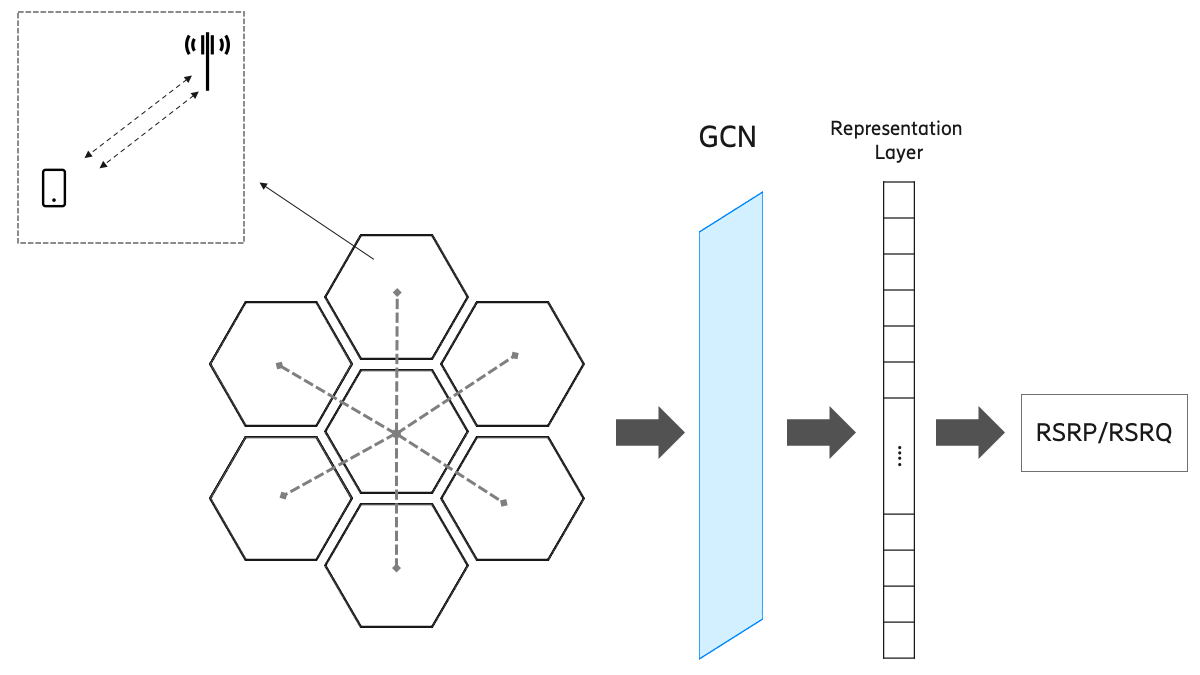}
    \caption{REMs estimation using GCN model.}
    \label{fig:method_overview}
\end{figure}

GNNs \cite{xia2021graph} are a class of machine learning models designed for learning and reasoning over graph-structured data. The network architecture for this task is Graph Conventional Networks (GCNs) \cite{kipf2017semisupervised}. The idea behind such models is to aggregate neighboring nodes’ content features while taking graph structures (edges/topology) into consideration. In our work, we represent a geographical region (e.g., city) using a set of tiles. Each tile is seen as a graph node. Figure \ref{fig:method_overview} depicts how coverage areas are represented as a graph. Links between coverage areas exist when the network nodes are geographically located in close proximity. 
Then, we use Graph Conventional Networks \cite{kipf2017semisupervised} to predict signal strength values characterized by RSRQ and RSRP. 

The model was trained in an unsupervised manner. The labeled samples from the dataset were used in the evaluation phase to quantify the performance of the prediction. 

\noindent\textbf{Regression Task:}
Given a temporal graph $(\mathcal{G}_{\text{C}}^t)_{t}$=$(\mathcal{V}^t, \mathcal{E}^t)_t$ as input with a time length of $T$, we aim to predict:
\[(\mathbf{y}^t)_{t=1}^{T} = \{y_1^t, \ldots, y_K^t\}_{t=1}^{T}\]

where $y_k^t$ denotes the predicted continuous value of signal quality.

\noindent\textbf{Classification Task:}
Given a temporal graph $(\mathcal{G}_{\text{C}}^t)_{t}$=$(\mathcal{V}^t, \mathcal{E}^t)_t$ as input with a time length of $T$, we aim to predict:
\[(\mathbf{y}^t)_{t=1}^{T} = \{y_1^t, \ldots, y_K^t\}_{t=1}^{T}\]
where $y_k^t \in \{0, 1, 3, 4\}$ denotes the quality of signal based on Table \ref{tab:rsrp_classification}.
\section{Experiments}
\label{section:experiments}

We compare our proposed method against several state-of-the-art models for multivariate time-series prediction and classification. The models are trained on the same features described in section 
\ref{section:method}:
\begin{itemize}
    \item \textbf{Model 1}: XGBoost \cite{chen2016xgboost}: an implementation of gradient-boosted decision trees designed for speed and performance, XGBoost is a powerful approach for regression and classification problems on tabular data.
    \item \textbf{Model 2}: FCN: fully connected neural network with 3 hidden layers of dimensions 256-512-128.
    \item \textbf{Model 3}: TabNet \cite{arik2021tabnet}: Attention-based neural network approach for tabular data, TabNet uses sequential attention to choose which features to process at each decision step, making it highly interpretable and capable of handling both sparse and dense features, as in our case of RSRP signal quality.
    \item \textbf{Model 4}: GCN \cite{kipf2017semisupervised}: Graph Convolutional Network. Since the sites in the network acquire topological attributes and are interconnected, the application of graph theory via GCNs becomes critical. These networks extend the convolutional concept to graph-structured data, allowing the model to learn a representation of each site by considering not just its own attributes but also those of its neighbors. Such information is aggregated via the signal measurements as well as the geospatial data. This is particularly useful for RSRP KPI prediction, as the signal quality at one site may be influenced by the signal qualities of neighboring sites. This leads to more accurate and robust models for network coverage estimation.
\end{itemize}

\subsection{Evaluation Metrics}

To assess the performance of our models, we divide our dataset into 20\% for testing and 80\% for training. We use well-established metrics: the $R^2$ score for regression tasks, and accuracy for classification tasks. 
The accuracy metric is an indicator of the model's overall classification performance, representing the ratio of correctly predicted observations to the total observations.
\begin{equation}
Accuracy = \frac{TP+TN}{TP+TN+FP+FN}
\end{equation}
where $tp$ stands for true positive, $fp$ stands for false positive, and $fn$ stands for false negative.

Our choice of loss functions is aligned with these metrics: cross-entropy for classification models and root-mean-squared error for regression models.

To ensure a comprehensive evaluation, the results, including accuracy and $R^2$ scores, are averaged across all H3 index areas. This averaging process provides a more robust and generalizable measure of model performance, taking into account the variability and specific characteristics of different geographical regions covered by each H3 index area.

\subsection{Regression Task}

In the regression task, the performance of the models was evaluated using the R2 score. As presented in Table \ref{tab:regression_task_results}, the GCN model outperformed the other models with an R2 score of 0.83, indicating a high level of predictive accuracy and suggesting that the GCN's structure is well-suited to capturing the topological and relational data inherent in the network's dataset. The FCN also showed strong performance with a score of 0.77, followed closely by TabNet with 0.74, which highlights the efficiency of attention mechanisms in tabular data prediction tasks. XGBoost, with a score of 0.62, lagged behind the neural network-based models, potentially due to its comparatively lower capacity to model complex non-linear interactions in the data. These results demonstrate the varying levels of efficacy of state-of-the-art models in handling the intricacies of regression tasks within multivariate time-series data.

\begin{table}[ht]
\centering
\begin{tabular}{|c|c|}
\hline \textbf{Model}         & \textbf{R2} \\ \hline 
XGBoost       &    0.62   \\  \hline
FCN           &    0.77   \\ \hline
TabNet        &    0.74   \\ \hline
GCN           &    0.83   \\ \hline  
\end{tabular}
\caption{The numerical results of the evaluated models (XGBoost, FCN, TabNet, GCN) for the regression task using the R2 score.}
\label{tab:regression_task_results}
\end{table}

\subsection{Classification Task}

For the classification task, we assess the accuracy of the same four models.
 Table \ref{tab:classification_task_results} shows the accuracy scores, where the GCN again demonstrates superior performance with an accuracy of 92\%. This suggests that the GCN's ability to leverage the relational structure of the data is particularly advantageous in classification tasks, allowing it to achieve a high degree of precision. Both the FCN and TabNet report a commendable accuracy of 74\%, indicating that these models are also effective at capturing the relevant patterns within the data for classification purposes. XGBoost, while still achieving a moderate level of accuracy at 63\%, appears less adept at this task compared to the more complex neural network-based approaches. These results underscore the importance of choosing a model that aligns well with the nature of the data and the specific requirements of the classification task at hand.
\begin{table}[ht]
\centering
\begin{tabular}{|c|c|}
\hline \textbf{Model}         & \textbf{Accuracy} \\ \hline 
XGBoost       &    0.63      \\  \hline
FCN           &    0.74      \\  \hline
TabNet        &    0.74      \\  \hline
GCN           &    0.92      \\ \hline  
\end{tabular}
\caption{The numerical results of the evaluated models (XGBoost, FCN, TabNet, GCN) for the classification task using the Accuracy score.}
\label{tab:classification_task_results}
\end{table}
\section{Conclusion}
\label{section:conclusion}

In this paper, we have presented an approach to impute and predict signal quality in geographical areas, which is fundamental to the creation of accurate and reliable Radio Environment Maps. The method provides an approach based on Graph Neural Networks for predicting signal quality maps in telecom networks using sparse signal strength measurements. 
The performance of GNN-based models showed better results (in both classification and regression tasks) as compared to models presented in other works (XGBoost, MLP, and TabNet). GNNs are capable of capturing the complex spatial relationships inherent in network topologies, thereby providing a more accurate representation of the inter-dependencies between network nodes. This approach not only enhances the accuracy of signal quality predictions but also offers insights into the spatial dynamics of network coverage.

To further improve the capabilities and applicability of GNN models for REMs, future work should prioritize the following tasks: 

- Investigating multimodal deep learning architectures. Our initial experiments using satellite imagery did not show any improvement. Many recent works showed promising results when incorporating environmental data. This may involve exploring the correlation between physical infrastructure and signal quality, enhancing the model’s ability to predict network performance in varying geographic and urban landscapes.

- Focusing on the refinement and optimization of the GCN model, enhancing its predictive accuracy and efficiency. This may involve exploring various neural network architectures, advanced training techniques, and hyperparameter optimization. 


\bibliographystyle{ieeetr}
\bibliography{bibliography}

\end{document}